\documentclass[aps,showpacs,pre,floatfix]{revtex4}
\usepackage{epsfig}
\usepackage{subfigure}

\begin{document}

\title{The memory kernel of velocity autocorrelation
function on a NiZr-liquid: theory and simulation}
\author{A.B. Mutiara}
\affiliation{Department of Informatics Engineering, Faculty of
Industrial Technology, Gunadarma University\\Jl. Margonda Raya No. 100, Depok 16424, Indonesia\\
E-mail:amutiara@staff.gunadarma.ac.id}


\begin{abstract}
\noindent Based on Sj\"{o}gren und Sjolander's
Mode-Coupling(MC)-Model, we have reformulated and calculated the
memory kernel (MK) of the velocity autocorrelation function
(VACF)on a Ni$_{0.2}$Zr$_{0.8}$-liquid. Reformulating means here
that we have constructed the memory kernel of VACF for our binary
system, instead of one for one atomic system of the Sj\"{o}gren
und Sjolander's model. The data required for the theoretical
calculations have been obtained from molecular dynamics (MD)
simulations. The theoretical results then are compared with those
directly obtained from computer simulation. We found, although it
exists a qualitative agreement between theoretical predictions and
simulation results, that quantitatively there is an deviations
between both results, especially for Zr-subsystem.

\end{abstract}
\maketitle
\section{Introduction}
Since last decade and until today, one important goal of modern
research in condensed matter is to find a good quantitative
description of the glassy dynamics in liquids. Theoretically there
is a established theory, the so-called mode-coupling theory (MCT),
that can describe and predict important parameters and quantities
in glassy systems qualitatively good. Work in the last decade has
provided evidence~\cite{vigo98} that MCT~\cite{gotze99} is able to
describe the slow dynamics of {\it fragile} liquids in the weakly
supercooled state. The recent work of Kob et.al.\cite{kob2000} has
also showed that theory able to accurately describe the
non-ergodicity parameters of simple as well as of network-forming
liquids. The one important statement of the MCT is that the
dynamics processes at long times is driven by a memory kernel
which is included in MCT's integro-differential equation.

In previously paper \cite{teimut2001} we have studied the memory
kernel of incoherent intermediate scattering function. Further,
we have compared the memory kernels $M^s_i(q,t)$ evaluated from
the MCT formula with the kernels $M^0_i(q,t)$ from inverting the
time evolution of the intermediate scattering functions. The
comparison shows encouraging agreement at 900 to 1100 K while
significant deviations are found at 1200 K. It is an open question
whether for this temperature an improved agreement between
calculated $M^s_i(q,t)$ and estimated $M^0_i(q,t)$ can be obtained
by inclusion of the coupling to transversal currents in the memory
kernel formula as provided, e.g., by Gudowski et.al.
\cite{gudowski}.

The main objective of paper is to give one of the answer of above
open question. But instead of the memory kernel of incoherent
intermediate scattering function, we investigated the memory
kernel $K(t)$ of the velocity autocorrelation function (VACF) for
temperature $T=1500$~K. By calculating the memory kernel we
included all coupling as proposed by Sj\"{o}gren und Sjolander's
Mode-Coupling(MC)-Model \cite{sjoesjol}. The calculated memory
function then is compared with the kernel from inverting the time
evolution of VACF.

Our paper is organized as follows: In Section \ref{SIM}, we
present the model and give some details of the computations.
Section \ref{THEO} gives  a brief discussion of some aspects of
the Sj\"{o}gren und Sjolander's Mode-Coupling(MC)-Model as used
here. Results of our MD-simulations and their analysis are then
presented and discussed in Section \ref{RD}.

\section{Sj\"{o}gren-Sj\"{o}lander's Mode Coupling
Model}\label{THEO} As we know, the VACF can be studied through the
formalism developed by Zwanzig and Mori \cite{zwanmori}, which is
based on a following integral equation:
\begin{equation}
\frac{d\psi(t)}{dt}= - \int^t_0 K(t')\psi(t-\tau) dt' \label{SS.0}
\end{equation}

In this paper the Sj\"{o}gren und Sjolander's (SS)
Mode-Coupling(MC)-Model \cite {sjoesjol} is used. In this model
memory kernel $K(t)$ of VACF $\psi (t)$ was proposed for a simple
liquid according to a combination of kinetic and mode coupling
concepts.

The basic idea of the model is that memory-kernel can be divided
into two term. The first term comes from the uncorrelated binray
collisions and the other from correlated collisions. The former is
related to the fast decay of kernel at short times. The latter is
expresses the mode-coupling term which incorporate more
sophisticated processes that appear at longer times. Moreover,
this term is based on the idea that the motion of a tagged
particle is influenced by constraints collectively imposed by its
neighboring particles. The memory kernel $K(t)$ of VACF take also
the following relation
\begin{equation}
K(t)=K_{B}(t)+K^{MC}(t)  \label{SS.1}
\end{equation}

According to SS-Model the mode-coupling effects take into account
four different coupling-term: density-density coupling, two
contributions from density-longitudinal current couplings, and
density-transverse current coupling (s. Sj\"{o}gren (1980/81)
\cite{sjo80,sjo81}, Gudowski et.al. (1993) \cite{gudowski},
Canales et.al. (1997) \cite{canales}).

To compute the memory kernel of VACF, we follow method used by
Sj\"{o}gren (1980/81), Gudowski et.al. (1993), Canales et.al.
(1997). If by these authors one atomic systems only were
investigated, here we investigate  a binary system. Also we have
to develop and construct the formulation of coupling terms for our
binary system.

Here are the results of our reformulation of $K(t)$ for our binary
system. The binary term is defined by \cite{balucani}
\begin{equation}
K_{B\alpha }(t)=\Omega _{E\alpha }^{2}\exp (-t^{2}/\tau _{\alpha }^{2}),
\label{SS.2}
\end{equation}
where $\Omega _{E\alpha }^{2}$ is the Einstein-frequency of $\alpha $%
-component \cite{bernu},
\begin{equation}
\Omega _{E\alpha }^{2}=\frac{1}{3m_{\alpha }}\sum\limits_{\beta }\int
g_{\alpha \beta }(r)\nabla ^{2}\phi _{\alpha \beta }(r)d^{3}r  \label{SS.3}
\end{equation}
and $\tau_{\alpha}$ can be determined by the following equation
\begin{equation}
\frac{\Omega _{E\alpha }^{2}}{\tau _{\alpha }^{2}}=-\frac{1}{2}\ddot{K}(0)
\label{SS.4}
\end{equation}

The coupling terms is expressed by
\begin{equation}
K_{\alpha }^{MK}(t)=K_{00\alpha }(t)+K_{01\alpha }(t)+K_{11\alpha
}(t)+K_{22\alpha }(t).  \label{SS.5}
\end{equation}
Here $K_{00\alpha }(t)$ describes the density-density coupling,
$K_{01\alpha }(t)$ and $K_{11\alpha }(t)$ the density-longitudinal
currents couplings, and $K_{22\alpha }(t)$ the density-transverse
current coupling.

The Laplace transforms of these terms can be written

\begin{equation}
\widetilde{K}_{00\alpha }(z)=\widetilde{R}_{00\alpha }(z),  \label{SS.6}
\end{equation}

\begin{equation}
\widetilde{K}_{01\alpha }(z)=\widetilde{R}_{01\alpha }(z)\left[ \widetilde{K}%
_{B\alpha }(z)+\widetilde{K_{\alpha }}(z)\right] ,  \label{SS.7}
\end{equation}

\begin{equation}
\widetilde{K}_{11\alpha }(z)=\widetilde{K}_{B\alpha }(z)\widetilde{R}%
_{11\alpha }(z)\widetilde{K}_{\alpha }(z),  \label{SS.8}
\end{equation}

\begin{equation}
\widetilde{K}_{22\alpha }=\left[ \widetilde{K}_{B\alpha }(z)+\widetilde{R}%
_{00\alpha }(z)+\widetilde{K}_{B\alpha }(z)\widetilde{R}_{01\alpha
}(z)\right] \widetilde{R}_{22\alpha }(z)\widetilde{K}_{\alpha }(z),
\label{SS.9}
\end{equation}
where $\widetilde{K}_{B\alpha }(z)$ und $\widetilde{K}_{\alpha
}(z)$ are repectively the Laplace transform of the binary term and
the total memory kernel. The Laplace transform of the total Memory
kernel is as follows

\begin{equation}
\widetilde{K}_{\alpha }(z)=\left[ \widetilde{K}_{B\alpha }(z)+\widetilde{R}%
_{00\alpha }(z)+\widetilde{K}_{B\alpha }(z)\widetilde{R}_{01\alpha
}(z)\right] /X  \label{SS.10}
\end{equation}
with
\begin{eqnarray}
X &=&1-\widetilde{R}_{01\alpha }(z)-\widetilde{K}_{B\alpha }(z)\widetilde{R}%
_{11\alpha }(z)-  \nonumber \\
&&\left[ \widetilde{K}_{B\alpha }(z)+\widetilde{R}_{00\alpha }(z)+\widetilde{%
K}_{B\alpha }(z)\widetilde{R}_{01\alpha }(z)\right] \widetilde{R}_{22\alpha
}(z)
\end{eqnarray}
where $\widetilde{R}_{ij\alpha }$ are the Laplace transform of
''recollision'' terms $R_{ij\alpha }(t)$. Assuming isotropic systems $%
R_{ij\alpha }(t)$ can be expressed as follows

\begin{equation}
R_{00\alpha }(t)=\frac{k_{B}T}{6\pi ^{2}m_{\alpha }\rho }\int\limits_{0}^{%
\infty }q^{4}(cF)_{\alpha }(q,t)\triangle F_{s\alpha }(q,t)dq,  \label{SS.11}
\end{equation}
with

\begin{eqnarray}
(cF)_{\alpha }(q,t) &=&\left[ c_{\alpha \alpha }(q)\right] ^{2}F_{\alpha
\alpha }(q,t)+  \nonumber \\
&&2c_{\alpha \alpha }(q)c_{\alpha \beta }(q)F_{\alpha \beta }(q,t)+\left[
c_{\alpha \beta }(q)\right] ^{2}F_{\alpha \alpha }(q,t);\text{ }\alpha \neq
\beta \qquad  \label{SS.12}
\end{eqnarray}

\begin{equation}
R_{01\alpha }(t)=-\frac{1}{6\pi ^{2}\Omega _{E\alpha }^{2}\rho }%
\int\limits_{0}^{\infty }q^{2}(LF^{\prime })_{\alpha }(q,t)\triangle
F_{s\alpha }(q,t)dq,
\end{equation}
with

\begin{eqnarray}
(LF^{\prime })_{\alpha }(q,t) &=&c_{\alpha \alpha }(q)\left[ \gamma
_{L\alpha \alpha }(q)+\frac{k_{B}Tq^{2}}{m_{\alpha }x_{\alpha }}c_{\alpha
\alpha }(q)\right] F_{\alpha \alpha }^{\prime }(q,t)+ \\
&&c_{\alpha \alpha }(q)\left[ \gamma _{L\alpha \beta }(q)+\frac{k_{B}Tq^{2}}{%
m_{\alpha }x_{\beta }}c_{\alpha \beta }(q)\right] F_{\alpha \beta }^{\prime
}(q,t)+  \nonumber \\
&&c_{\alpha \beta }(q)\left[ \gamma _{L\alpha \alpha }(q)+\frac{k_{B}Tq^{2}}{%
m_{\alpha }x_{\alpha }}c_{\alpha \alpha }(q)\right] F_{\alpha \beta
}^{\prime }(q,t)+  \nonumber \\
&&c_{\alpha \beta }(q)\left[ \gamma _{L\alpha \beta }(q)+\frac{k_{B}Tq^{2}}{%
m_{\alpha }x_{\beta }}c_{\alpha \beta }(q)\right] F_{\beta \beta }^{\prime
}(q,t);\text{ }\alpha \neq \beta  \nonumber
\end{eqnarray}

\begin{equation}
R_{11\alpha }(t)=-\frac{1}{6\pi ^{2}\Omega _{E\alpha }^{4}\rho }%
\int\limits_{0}^{\infty }q^{2}(BC_{L})_{\alpha }(q,t)\triangle F_{s\alpha
}(q,t)dq,
\end{equation}
with
\begin{eqnarray}
(BC_{L})_{\alpha }(q,t) &=&\left[ \gamma _{L\alpha \alpha }(q)+\frac{%
k_{B}Tq^{2}}{m_{\alpha }x_{\alpha }}c_{\alpha \alpha }(q)\right]
^{2}C_{L\alpha \alpha }(q,t)+  \nonumber \\
&&2\left[ \gamma _{L\alpha \alpha }(q)+\frac{k_{B}Tq^{2}}{m_{\alpha
}x_{\alpha }}c_{\alpha \alpha }(q)\right] \left[ \gamma _{L\alpha \beta }(q)+%
\frac{k_{B}Tq^{2}}{m_{\alpha }x_{\beta }}c_{\alpha \beta }(q)\right]
C_{L\alpha \beta }(q,t)+  \nonumber \\
&&\left[ \gamma _{L\alpha \beta }(q)+\frac{k_{B}Tq^{2}}{m_{\alpha }x_{\beta }%
}c_{\alpha \beta }(q)\right] ^{2}C_{L_{\beta \beta }}(q,t);\text{ }\alpha
\neq \beta
\end{eqnarray}

\begin{equation}
R_{22\alpha }(t)=-\frac{1}{6\pi ^{2}\Omega _{E\alpha }^{4}\rho }%
\int\limits_{0}^{\infty }q^{2}(TC_{T})_{\alpha }(q,t)\triangle F_{s\alpha
}(q,t)dq,
\end{equation}
with

\begin{eqnarray}
(TC_{T})_{\alpha }(q,t) &=&\left[ \gamma _{T\alpha \alpha }(q)\right]
^{2}C_{T\alpha \alpha }(q,t)+  \nonumber \\
&&2\gamma _{T\alpha \alpha }(q)\gamma _{T\alpha \beta }(q)C_{T\alpha \beta
}(q,t)+\left[ \gamma _{T\alpha \beta }(q)\right] ^{2}C_{T\alpha \alpha }(q,t)
\end{eqnarray}
Here $c_{\alpha \alpha }(q)$ is the Fourier transform of $\alpha
\alpha -$~direct correlation function, $\gamma _{L\alpha \alpha }(q)$ and $%
\gamma _{T\alpha \alpha }(q)$ are $q$~-dependent quantities, which
are defined in Balucani \cite{balucani}. $C_{T\alpha \alpha
}(q,t)$ und $C_{L\alpha \alpha }(q,t)$ are the transverse and
longitudinal current correlation functions of the $\alpha \alpha
$-~parts. $\triangle
F_{s\alpha }(q,t)$ is the difference between $F_{s\alpha }(q,t)$ and $%
F_{0\alpha }(q,t)$, where $F_{0\alpha }(q,t)=\exp
(-(k_{B}T/2m_{\alpha })q^{2}t^{2})$ expresses the
free-particle-form of intermediate incoherent scattering function.

\section{SIMULATIONS \label{SIM}}
The simulations are carried out as state-of-the-art
isothermal-isobaric ($N,T,p$) calculations. The Newtonian
equations of $N=$ 648 atoms (130 Ni and 518 Zr) are numerically
integrated by a fifth order predictor-corrector algorithm with
time step $\Delta t$ = 2.5 10$^{-15}$s in a cubic volume with
periodic boundary conditions and variable box length L. With
regard to the electron theoretical description of the interatomic
potentials in transition metal alloys by Hausleitner and Hafner
\cite{haushafner}, we model the interatomic couplings as in
\cite{tei92} by a volume dependent electron-gas term $E_{vol}(V)$
and pair potentials $\phi(r)$ adapted to the equilibrium distance,
depth, width, and zero of the Hausleitner-Hafner potentials
\cite{haushafner} for Ni$_{20}$Zr$_{80}$ \cite{thesis}. For this
model, simulations were started through heating a starting
configuration up to 2000~K which leads to a homogeneous liquid
state. The system then is cooled continuously to various annealing
temperatures with cooling rate $-\partial_tT$ = 1.5 10$^{12}$~K/s.
Afterwards the obtained configurations at various annealing
temperatures (here 1500-800 K) are relaxed by carrying out
additional isothermal annealing runs. Finally the time evolution
of these relaxed configurations is modelled and analyzed. More
details of the simulations are given in \cite{thesis}.

\section{Result and Discussion}\label{RD}
We have calculated the total memory kernel $K_{\alpha }(t)$ of VACF in range $%
0,024\leq q\leq 4,796$ \AA $^{-1}$. The input data are all
correlation functions that are analyzed from MD data.
For simplifying we have only calculated the memory kernel at $%
T=1500$~K. The reason for this is, because (i) according to some
authors, except binary term and density-density coupling, the
other couplings could be neglected by lower temperature, i.e., the
contribution of the other couplings are very small, and more
smaller than the contribution of the binary term and
density-density coupling \cite{bgs}, (ii) after our experience
there is a coupling contribution which decays slowly at lower
temperature. These slowly decays make us technically a problem by
the Laplace transform of couplings.

\begin{figure}[htbp]
\begin{center}
\leavevmode
\includegraphics[width=3.in,height=2.1in]{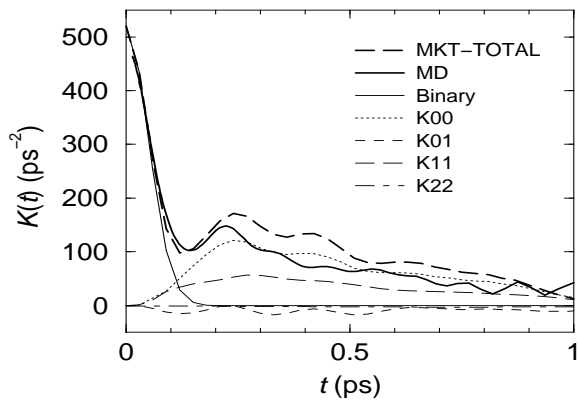}
\caption{Ni - $K(t)/K(0)$. } \label{MVNINEUFT}
\end{center}
\end{figure}

\begin{figure}[htbp]
\begin{center}
\leavevmode
\includegraphics[width=3in,height=2.1in]{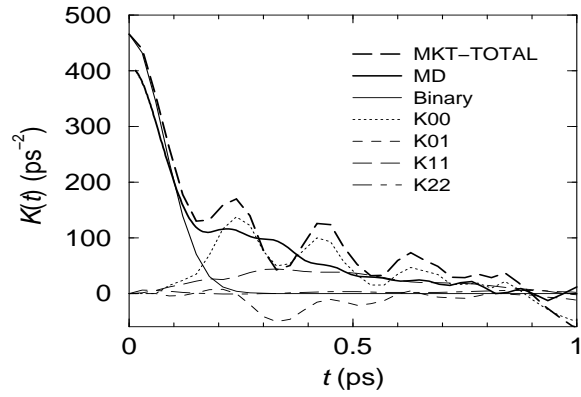}
\caption{Zr - $K(t)/K(0)$. } \label{MVZRNEUFT}
\end{center}
\end{figure}

\begin{figure}[htbp]
\centering \leavevmode
\includegraphics[width=3in,height=2.in]{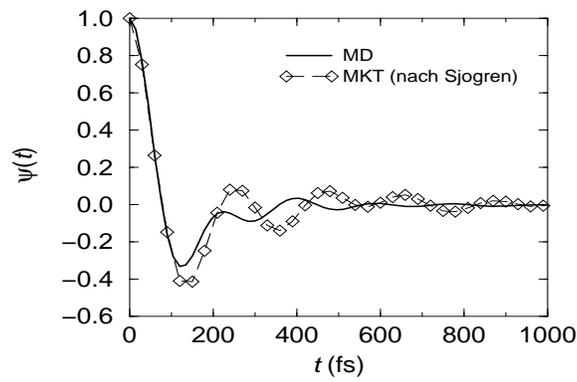}
\caption{Ni - $\psi (t)$. } \label{VMDMC15NI}
\end{figure}

\begin{figure}[htbp]
\centering \leavevmode
\includegraphics[width=3in,height=2in]{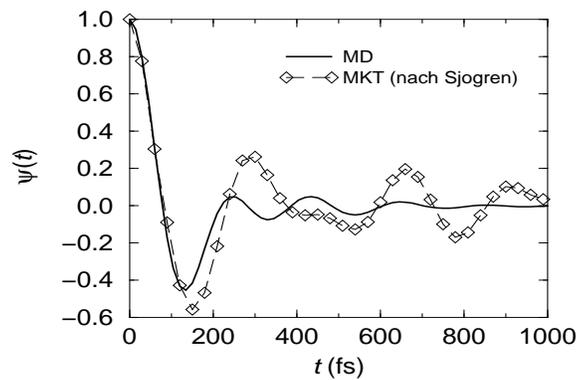}
\caption{Zr - $\psi (t)$. } \label{VMDMC15ZR}
\end{figure}

The figures (\ref{MVNINEUFT}) and (\ref{MVZRNEUFT}) shows our
results of the memory kernel $K_{\alpha }(t)$ of VACF. MD-results
can be computed over the equation (\ref{SS.0}). From figures we can
see that the results of SS-model do not have a good agreement with
MD-results, escpecially those for Zr-atom. By Zr-atom the run of the
memory kernel $K_{\alpha }(t)$ shows some of oscillations that by
MD-results do not appear. These oscilations are primaly from the
density-density coupling. According to eq.(\ref{SS.12}) the
intermediate coherent scattering functions $\mathbf{F}(q,t)$ are
responsible for those oscilations. The deviations between SS-models
results and MD-results can also be shown by results in computing
$\psi (t)$, e.g., on figures (\ref{VMDMC15NI}) and
(\ref{VMDMC15ZR}).

As shown in figures (\ref{MVNINEUFT}) and (\ref{MVZRNEUFT}) the
most significant contributions come from the density-density
coupling, the second  from the density-longitudinal current
density coupling, and the least from the density-transverse
current coupling.


\section{Concluding Remarks}
We have calculated the memory kernel of VACF based on SS-model.
The parameters used by calculating have produced from
MD-simulation. We have compared the SS-model's results and
MD-results. Both results are shown a qualitative agreement,
although quantitatively there is a deviations between both
results, especially for Zr-subsystem. These results are agreed
with the generall Mode Coupling Theory's predictions
\cite{gotze99}, namely that the most significant contributions are
resulted from the density-density coupling. This coupling always
is dominant when the temperature of system is lowered, and
especially when the temperature of system is near the critical
glass temperature $T_c$.

\end{document}